\documentclass[lettersize,journal,oneside]{IEEEtran}
\IEEEoverridecommandlockouts
\usepackage{cite}
\usepackage{amsmath,amssymb,amsfonts}
\usepackage{algorithmic}
\usepackage{graphicx}
\usepackage{textcomp}
\usepackage{xcolor}
\usepackage{subfigure}
\usepackage{fancyhdr}
\usepackage{marvosym}
\usepackage{booktabs}     
\usepackage{caption}      
\usepackage{hyperref}
\def\BibTeX{{\rm B\kern-.05em{\sc i\kern-.025em b}\kern-.08em
    T\kern-.1667em\lower.7ex\hbox{E}\kern-.125emX}}
    
\begin{document}
\title{Adaptive Implicit-Based Deep Learning Channel Estimation for 6G Communications\\}
\author{Zhen Qiao, Jiang Xue, Junkai Zhang, Guanzhang Liu, Xiaoqin Ma, Runhua Li, Faheem A. Khan, John S. Thompson, and Zongben Xu%
\thanks{Zhen Qiao, Junkai Zhang (corresponding author), Guanzhang Liu, Xiaoqin Ma, Runhua Li, and Zongben Xu are with School of Mathematics and Statistics, Xi'an Jiaotong University.}
\thanks{Jiang Xue is with School of Mathematics and Statistics, Xi'an Jiaotong Univeristy, and also with Peng Cheng Laboratory, and also with Pazhou
Laboratory (Huangpu).}
\thanks{Faheem A. Khan is with School of Computing and Engineering, University of Huddersfield.}
\thanks{John S. Thompson is with School of Engineering, The University of Edinburgh.}
}
\maketitle

\section*{Abstract}
With the widespread deployment of fifth-generation (5G) wireless networks, research on sixth-generation (6G) technology is gaining momentum.
Artificial Intelligence (AI) is anticipated to play a significant role in 6G, particularly through integration with the physical layer for tasks such as channel estimation.
Considering resource limitations in real systems, the AI algorithm should be designed to have the ability to balance the accuracy and resource consumption according to the scenarios dynamically.
However, conventional explicit multilayer-stacked Deep Learning (DL) models struggle to adapt due to their heavy reliance on the structure of deep neural networks.
This article proposes an adaptive Implicit-layer DL Channel Estimation Network (ICENet) with a lightweight framework for vehicle-to-everything communications.
This novel approach balances computational complexity and channel estimation accuracy by dynamically adjusting computational resources based on input data conditions, such as channel quality.
Unlike explicit multilayer-stacked DL-based channel estimation models, ICENet offers a flexible framework, where specific requirements can be achieved by adaptively changing the number of iterations of the iterative layer.
Meanwhile, ICENet requires less memory while maintaining high performance. The article concludes by highlighting open research challenges and promising future research directions.

\section{Introduction}
\label{introduction}
\noindent 
Artificial intelligence (AI), particularly Deep Learning (DL), is leading innovation across multiple fields such as computer vision, natural language processing, and gaming. The integration of AI with sixth-generation (6G) mobile communications technologies promises to fulfill the vision of ``ubiquitous intelligence and connectivity" \cite{9770094}. Vehicle-to-everything (V2X) communications have been highlighted as a key enabling technology to satisfy the demands of the future intelligent transportation systems, which will benefit from 6G technologies designed to support a wide range of highly mobile users \cite{9923616}. However, the presence of non-stationary channels in V2X communications affects the performance in high mobility scenarios, and an in-depth study of these channels and their estimation methods becomes essential. 

Orthogonal frequency-division multiplexing (OFDM) is a widely adopted multi-carrier technique in wireless communications. However, the fast movement of the receivers drastically alters the scattering conditions of the channel, causing the channel response to exhibit a birth-death process. These rapid channel variations require channel models that capture the stationary intervals of the channel accurately \cite{9786750}. Traditional channel modeling often employs several assumptions and approximations, such as linearity and time-invariance, to simplify their analysis in wireless communications \cite{9023918}. While these simplifications reduce computational complexity, they compromise modeling precision.

Meanwhile, the number of channel coefficients that need to be estimated increases dramatically for rapidly changing channels in high-speed V2X scenarios, further increasing the complexity of channel estimation. To cope with this challenge, channel estimation techniques usually consider the underlying properties of the channel to simplify the model\cite{BEM}, such as reducing the number of parameters to be estimated or assuming that the channel varies linearly over multiple OFDM symbol periods. However, the effectiveness of these approaches relies on accurate modeling of the channel variations. Therefore, high mobility scenarios impose higher requirements on the accuracy and real-time performance of channel estimation algorithms. It is imperative to explore more advanced methods that can adapt to the rapid changes in the channel. DL, as a powerful branch of AI, offers new perspectives and solutions for addressing such complex and dynamic problems.

Recently, DL techniques have been used to design key modules in mobile communications systems, including channel estimation, prediction, feedback, etc.\cite{8715338,hu2020deep, 10075639}. However, the application of DL in mobile communications faces challenges in minimizing the computation and communication cost, compressing training data, and reducing computational load\cite{9770094}. In addition, most of the existing AI algorithms are designed for specific problems and the computational complexity is fixed \cite{soltani2019deep}. Nevertheless, hardware resources impose limitations on algorithm complexity in practical applications. Therefore, meeting the performance demands of 6G communication requires an adaptive DL channel estimation algorithm that balances computational complexity and accuracy.

Deep neural networks (DNNs) are typically constructed by stacking multiple layers/blocks to extract features from input data, ultimately producing accurate and explicit outputs. However, these explicit DNNs experience several challenges, such as
i) the need to choose the appropriate number of layers/blocks to avoid vanishing and exploding gradients in training;
ii) limited computational resources that cannot support very large DNNs; 
iii) the stacked structure of different layers/blocks makes DNNs fixed and hard to adjust based on input.

Against this backdrop, implicit DL architectures have gained increasing attention in academia. Unlike traditional explicit architectures that sequentially stack deterministic computational modules, implicit DL networks define input-output relationships as an implicitly joint function \cite{bai2022equilibrium}. Mathematically, when the output of an implicit DL network satisfies certain constraints, this implicit relationship can be formulated as fixed-point equations, differential equations, or other forms, offering a new paradigm for DL research.

In this article, we propose a lightweight adaptive Implicit-layer DL Channel Estimation Network (ICENet) for 6G high-mobility V2X scenarios. Firstly, ICENet adopts a lightweight single implicit-layer/block DL architecture that simulates the depth of conventional explicit DNNs through iteration. This design maintains a fixed and low memory consumption, offering an efficient solution for channel estimation in resource-constrained terminal environments. Secondly, by incorporating adaptive stopping criteria, ICENet dynamically adjusts computational depth based on the noise level of the input CSI. This enables a trade-off between computational complexity and estimation accuracy, addressing inefficiencies caused by fixed structures in manually designed models. Finally, the numerical results show that ICENet achieves superior performance and adaptability, establishing it as a promising and efficient channel estimation method that opens up a new path for wireless communication in the 6G era.

\section{Channel Estimation in V2X Communications}
Channel estimation aims to accurately reconstruct the entire channel response with limited channel state information (CSI). Traditional channel estimation involves two main categories: reference-based and blind estimation. The former leverages known signals (training symbols or pilots) to improve accuracy and performance, while the latter relies on statistical properties of received data requiring higher computational complexity. The fully blind and semi-blind channel estimation methods usually require the channel to remain stable over multiple symbols. In contrast, pilot-assisted channel estimation is widely used in OFDM systems, typically using techniques such as least squares (LS), minimum mean square error (MMSE) and linear MMSE (LMMSE) \cite{hu2020deep}.

However, the channel in high-speed mobile scenarios cannot be simply regarded as wide-sense stationary, which poses a challenge for the aforementioned traditional channel estimation methods. We give an example of time-frequency domain high-speed V2X channel characterization for the urban-macrocell (UMa) non-line-of-sight (NLoS) scenario in Fig.~\ref{fig: Qua}, generated by QUAsi Deterministic RadIo channel GenerAtor (QuaDRiGa), which is used to simulate the real-world V2X channel in this article \cite{quadriga}.
To cope with this rapidly changing channel, current channel estimation research can be analyzed in depth from the following three perspectives, which intersect with each other:

\begin{figure}
    \centering
    \includegraphics[width=0.7\columnwidth]{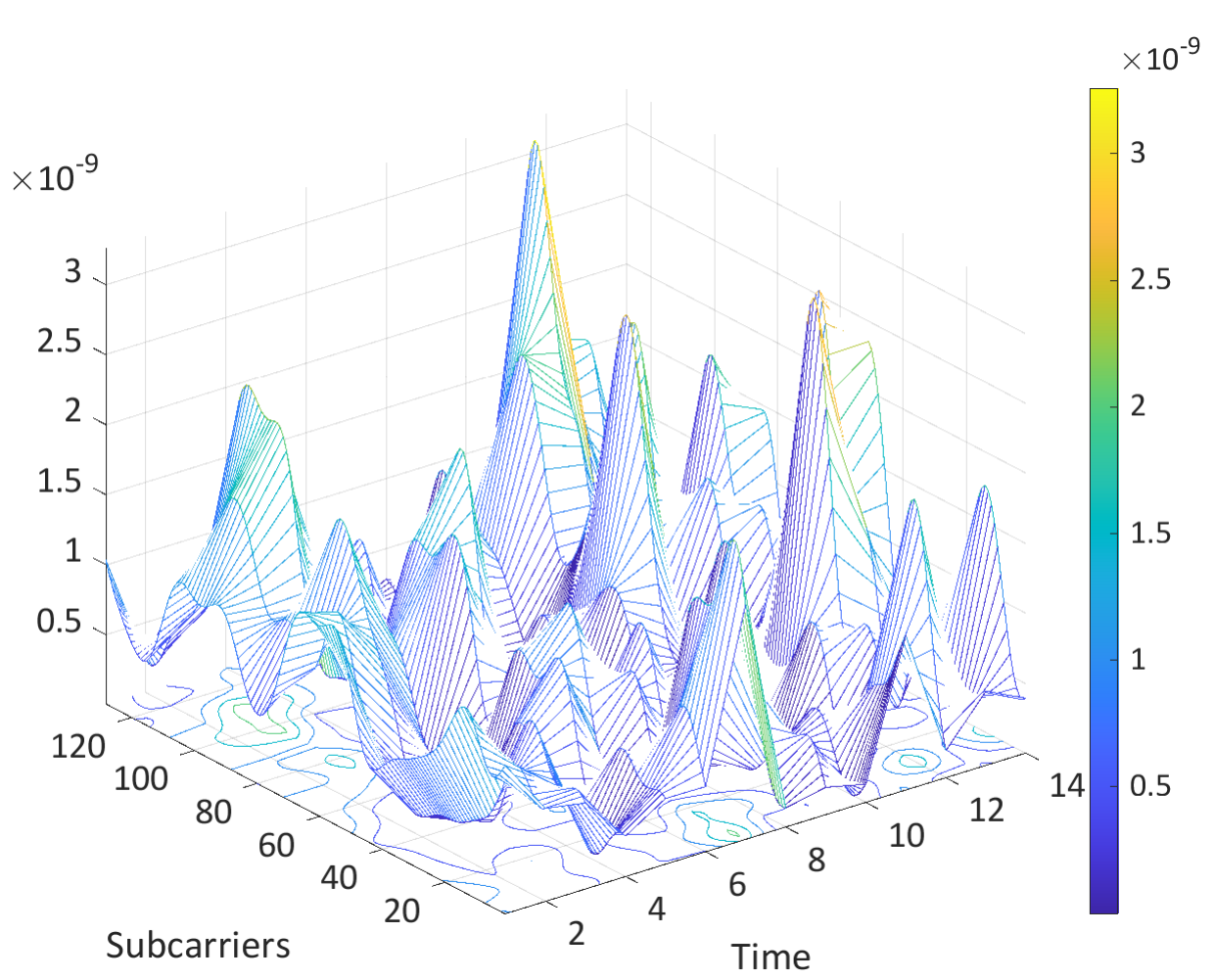}
    \caption{Time-frequency Domain V2X Channel Characterization for UMa-NLoS Scenarios Simulated by QuaDRiGa.}
    \label{fig: Qua}
\end{figure}

\begin{itemize}
    \item \textit{Interpolation Perspective:}
    {Conventional multiple-input multiple-output (MIMO) OFDM channel estimation techniques are typically based on the assumption of the high correlation between channel responses of adjacent OFDM symbols.
    These methods use various interpolation techniques to transform the channel estimation into an interpolation problem.
    }
    However, its effectiveness often relies on accurate modeling of channel variations and can be significantly distorted for high Doppler scenarios.
    
    \item \textit{Channel Model Perspective:}
    The basis expansion model (BEM) effectively models time-varying wireless channels as linear combinations of weighted basis functions to capture multi-path and Doppler effects, which is good for time-frequency dual-selective fading channels \cite{BEM}. However, the frequent parameter updates for accurately measuring the number of multi-path components and the delay in capturing the rapid channel changes further complicate the implementation. Additionally, the accuracy of BEM models heavily depends on the choice of basis functions, and they have high computational complexity.

    \item \textit{Modulation Perspective:} 
    In high-speed mobile scenarios, the Doppler shift may destroy the orthogonality between OFDM sub-carriers. Recently, orthogonal time-frequency space has been introduced as an innovative 2-dimensional (2D) modulation scheme to convert time-varying channels into static channels in the Delay-Doppler (DD) domain. Meanwhile, it can be combined with compressed sensing (CS) algorithms to simplify the estimation process by exploiting the channel sparsity in the DD domain. However, the high computational complexity of CS and the degree of sparsity of the DD channel may limit its effectiveness in practical applications.

\end{itemize}

Given the limited performance (e.g., LS) and high computational complexity (e.g., MMSE) of traditional algorithms in wireless communications, introducing AI technology into the physical layer (PHY) of wireless communications creates new possibilities for developing channel estimation algorithms. By utilizing the powerful computational efficiency, and self-learning capabilities of AI, computational complexity can be significantly reduced without sacrificing accuracy. Currently, DL techniques for channel estimation are divided into two main categories: data-driven approaches and data-model dual-driven approaches, which are reviewed below.

\begin{itemize}
\item \textit{Data-Driven Channel Estimation:}
Data-driven DL channel estimation techniques discard the reliance on traditional apriori knowledge or theoretical models. Instead, they leverage powerful data processing capabilities of AI to deeply mine and analyze high-quality input data \cite{hu2020deep}. In the early stage, researchers attempted to design the entire OFDM receiver with the DNNs rather than design each functional module of the receiver (e.g., channel estimation and signal detection module) separately. However, the DNNs become unexplainable due to the lack of expert knowledge in wireless communications. Currently, research focuses on using AI to optimize individual functional modules of the PHY, especially channel estimation. To this end, some researchers have drawn inspiration from advanced image processing techniques, such as AI-based super-resolution methods \cite{soltani2019deep}.

\item \textit{Data-Model Dual-Driven Channel Estimation:}
The data-model dual-driven approaches enable designing a DNN topology with theoretical foundations, making the network structure explainable and predictable \cite{xusun}. In channel estimation, data-model dual-driven approaches combine wireless communications knowledge with DL for DNNs with multiple learnable parameters. With the increasing number of antennas and larger bandwidths available at higher frequencies, DL is combined with CS to exploit the sparsity of millimeter-wave massive MIMO channels in the delay and angular domains and treat channel estimation as a sparse signal recovery problem. Consequently, some iterative channel estimation algorithms and their DL-based extensions for specific sparse structures such as orthogonal matching pursuit, approximate message passing\cite{8715338}, and iterative shrinkage-threshold algorithms are proposed\cite{10075639}. However, the performance of model-driven DL channel estimation is highly dependent on the quality of the model.
\end{itemize}

Existing DL-based channel estimation technologies have made significant progress, but their architecture relying on multilayer stacking is not flexible enough for practical applications. In addition, the large number of parameters leads to a high demand for computational resources, which limits its deployment on user equipments (UEs). Therefore, developing a lightweight DNN that can obtain a trade-off between accuracy and complexity is crucial for efficient channel estimation at UEs.

\section{Explicit Multilayer-Stacked Vs Implicit Equilibrium DNN Architectures}

\subsection{Explicit Multilayer-Stacked DNNs}\label{3a}
Recently, there has been a surge in the application of DL-based schemes across various fields, primarily due to their ability to identify and extract hidden complex structures and potential relationships within data. Most existing DL approaches rely on stacking existing layers/blocks or creating new ones to build network architectures for specific tasks.  DNN models such as Visual Geometry Group and Network in Network \cite{zhang2023dive} use various connection strategies to extract data features, often including typically integrating multiple non-linear transformations and may incorporate techniques such as normalization and dropout for performance optimization. These aforementioned explicit models view the deep network as a large computational graph, processing input through multiple layers to produce the desired output. Undoubtedly, layer stacking captures complex nonlinear relationships by extracting features at various levels to construct large-scale networks. Accompanying this, there are also several challenges:

\begin{itemize}
    \item \textit{Network Depth as Hyperparameter:} 
    Expanding network depth boosts model expressiveness but increases training complexity and computational costs concurrently. Furthermore, excessive depth may induce challenges such as gradient vanishing and overfitting.
    \item \textit{Large Memory Overhead:} 
    The gradient backpropagation algorithm is widely employed in training explicit networks. However, as the depth increases, more memory is required to store intermediate data. This can lead to memory resource constraints, thus limiting network scalability and training efficiency.
    \item \textit{Fixed MultiLayer-Stacked Structure:} 
    Current layer-stacked DNN architectures are often fixed and it is difficult to adjust structure based on input complexity dynamically unless using mechanisms like layer-skipping or early exiting. However, these mechanisms increase the complexity of network design. In other words, the layer-stacked DNN architectures introduce less flexibility which makes it difficult to realize the trade-off between complexity and accuracy.
\end{itemize}

An alternative DNN architecture, referred to as implicit equilibrium DNN, can alleviate challenges in existing explicit multilayer-stacked DNNs while maintaining similar performance. This alternative solution is presented in the next subsection.

\subsection{Implicit Equilibrium DNNs}
In real-world deployments, algorithm complexity should be kept within a manageable level due to hardware resource limitation. Additionally, the finite storage capacity of devices necessitates lightweight models that can be efficiently stored and fast deployed. Although building models to handle complex mapping relationships by stacking layers is still the dominant approach, implicit equilibrium DNN models bring an innovative framework to network design \cite{bai2022equilibrium}. This implicit network does not need to construct a large computational graph to form a specific mapping relationship between inputs and outputs.

The hidden layer of most existing explicit multilayer-stacked models can converge to a certain fixed point. Therefore, implicit equilibrium models convert the traditional explicit layer-stacked DNN into a network, whose weight and bias of each layer are the same, and represent the output by computing the fixed point. In this way, the implicit equilibrium models can be treated as a single layer/block DNN with an iteration process on it.

Specifically, given input $\mathbf{x}$ and a nonlinear transformation (often parameterized) $f_{\theta}(\cdot,\mathbf{x})$, the output is directly computed as the fixed point of a dynamic system. This computation can be achieved through any black-box root-finding method with minimal overhead, leading to the emergence of ``infinite-layer" equilibrium feature states. This solution corresponds to the last hidden layer value of the DNNs. Indeed, the emergence of implicit layers provides an alternative perspective for constructing DL architectures and brings the following advantages that explicit architecture lacks.

\begin{figure}
    \centering
    \includegraphics[width=\columnwidth]{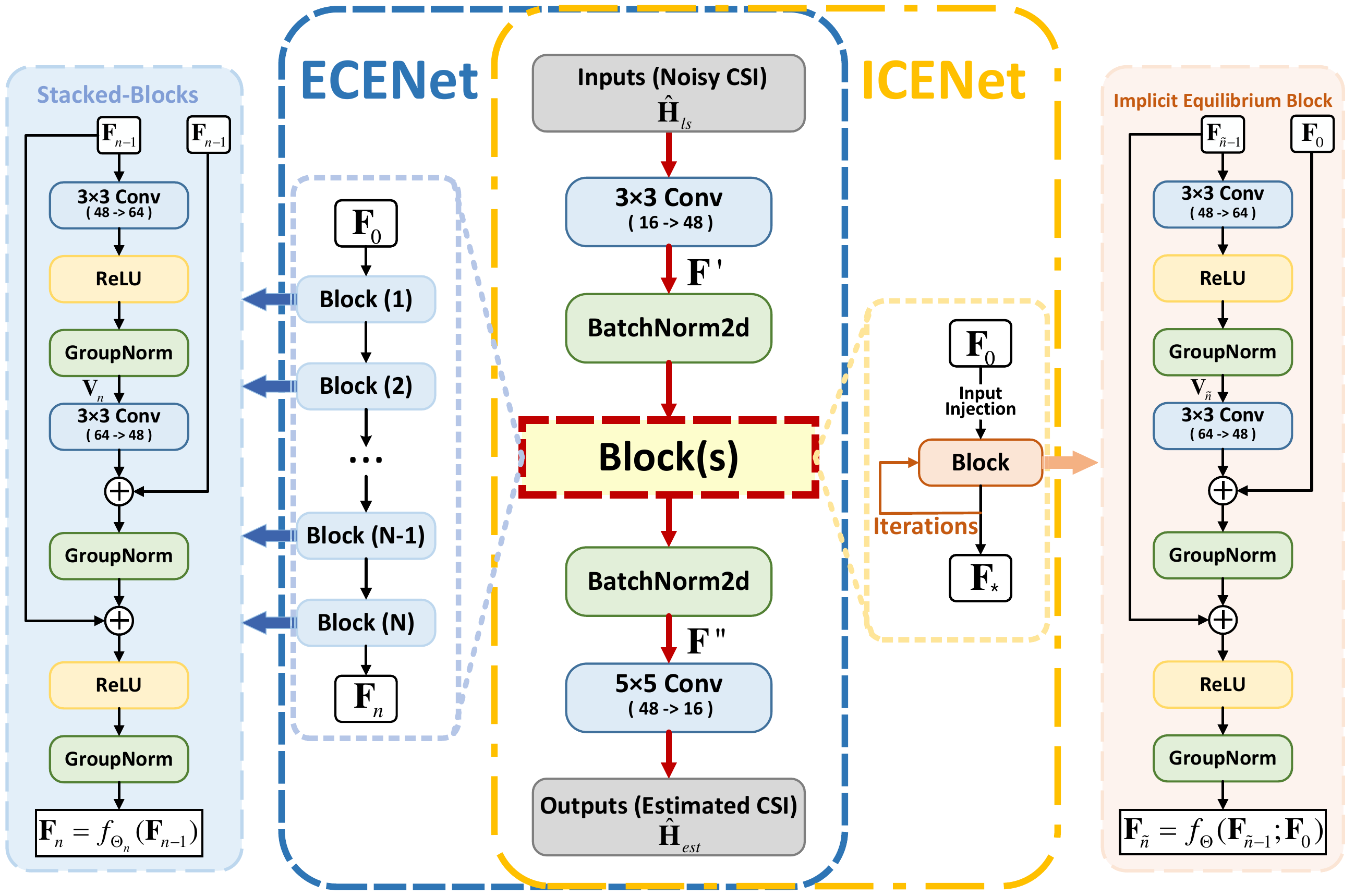}
    \caption{Comparsion of Explicit Multilayer-Stacked and Implicit Equilibrium DL Channel Estimation Architectures.}
    \label{fig: Conv}
\end{figure}

\begin{itemize}
    \item \textit{Layer-Free Structure:}
    The implicit network requires only one layer of $f_{\theta}(\cdot,\mathbf{x})$ in the entire DNN architecture without losing performance compared with the explicit architecture. This structure avoids the gradient vanishing problem issue to some extent and enhances the flexibility of DNN design.
    \item \textit{Constant and Small Memory Overhead:} 
    The training memory complexity of an explicit network with $L$-stacked layers is $O(L)$, while the implicit one consumes only $O(1)$ memory footprint, which allows the implicit network to support larger models with limited hardware resources.
    \item \textit{Decoupling:} 
    The implicit DNN approach decouples the representation capability of the model, controlled by the internal structure of $f_{\theta}(\cdot,\mathbf{x})$, and the actual computation, determined by the solver.
    Additionally, direct differentiation, via Implicit Function Theorem \cite{bai2022equilibrium}, of fixed point without explicit knowledge of the forward computation decouples forward and backward propagation.
\end{itemize}

\section{Adaptive Implicit DL-Based Channel Estimation Architecture}
In this section, we present adaptive implicit DL-based channel estimation architecture, termed ICENet, in detail from several perspectives.

\subsection{System Model}
We consider the MIMO-OFDM uplink channel estimation problem. The base station is equipped with 8 received antennas to serve 2 UEs, each with a single transmit antenna moving at their speeds, and the OFDM system has 128 subcarriers. Assuming that each OFDM frame has 14 OFDM symbols, we insert the pilot at the 2nd and 11th symbols of the frame. By exploring the orthogonality of subcarriers, we only insert the pilot every 2 subcarriers along all subcarriers to convert the MIMO channel estimation problem into two independent single-input multiple-output (SIMO) (i.e., $1\times8$ SIMO) channel estimation problems \cite{10310231}. We aim to design the channel estimation method to efficiently map from the estimated channel derived by pilots to the estimated full CSI.

\subsection{Proposed ICENet Model}
Most current AI-based channel estimation algorithms have fixed network structures. For example, we develop an Explicit Multilayer-stacked Channel Estimation Network (ECENet), as shown in the blue box on the left-hand side of Fig.~\ref{fig: Conv}, that stacks multiple $\operatorname{Blocks}$ with the same architecture but different parameters. As described in Section~\ref{3a}, the computational complexity and memory requirements of most explicit DNNs (e.g., ECENet) increase with network depth. Furthermore, ECENet faces challenges in adaptively adjusting its structure based on input quality, particularly for CSI, which is affected by varying noise levels. Building on this, we introduce the implicit DNN framework, called ICENet, shown in the yellow box on the right-hand side of Fig.~\ref{fig: Conv}, which simplifies the ECENet with $n$-stacked $\operatorname{Blocks}$ into a single-$\operatorname{Block}$ structure. This can be viewed as an infinitely deep, weight-tied DNN. Meanwhile, the memory cost of this framework is fixed and relatively small, enabling dynamic adjustment depth based on input CSI quality, making ICENet particularly well-suited for resource-constrained scenarios. Notably, in the ECENet and ICENet architectures, shallow layers employ small convolutional kernels to capture rapid fluctuations in local channel details and prevent overfitting to global data. Larger kernels are used last to integrate these local channel features.

\subsubsection{ICENet}
We introduce a lightweight design for the ICENet-implicit equilibrium block (IEB) (see right-hand side of Fig.~\ref{fig: Conv}), with the hyper-parameters of the DL module and activation functions labeled in the figure for clarity. ICENet eventually determines the fixed point $\mathbf{F}_{\star}$ based on the nonlinear transformation $f_{\theta}(\cdot)$ and the input $\mathbf{F}_{0}$. Notably, $\mathbf{F}_{\star}$ also represents the equilibrium state of the hidden layers within ICENet. ICENet dynamically adjusts the iteration count of its ICENet-IEB based on the input channel conditions, i.e., analogous to the depth in traditional explicit DNNs. This adaptability allows ICENet to use fewer iterations for simple channels and more for challenging ones, effectively managing computational costs across different scenarios. Furthermore, we establish a relationship between the output accuracy in ICENet-IEB and the normalized mean square error (NMSE) of channel estimation, where NMSE can evaluate the performance of different models or algorithms on a common scale. In addition, the performance of ICENet also hinges on the design of the ICENet-IEB.

As mentioned before, the final output of ICENet-IEB is the fixed point (i.e., the equilibrium point) itself. Specifically, this fixed point is determined by iterative root-finding algorithms during forward propagation, such as the Anderson acceleration (AA) method \cite{anderson1965iterative} employed in this study to achieve efficient convergence. Notably, ICENet decouples the forward and backward propagation processes, requiring storage only for $\mathbf{F}_{\star}$, $f_{\theta}(\cdot, \mathbf{F}_{0})$, and the input $\mathbf{F}_{0}$. In contrast, explicit DL-based channel estimation architectures such as ECENet must store temporary computation caches (e.g., intermediate activation outputs) necessary for gradient calculations, substantially increasing memory requirements during training. More importantly, by simply modifying $f_{\theta}(\cdot)$, the framework can be extended to construct more sophisticated networks for addressing channel estimation challenges in future communication scenarios.

\begin{figure}
    \centering
    \includegraphics[width=0.7\columnwidth]{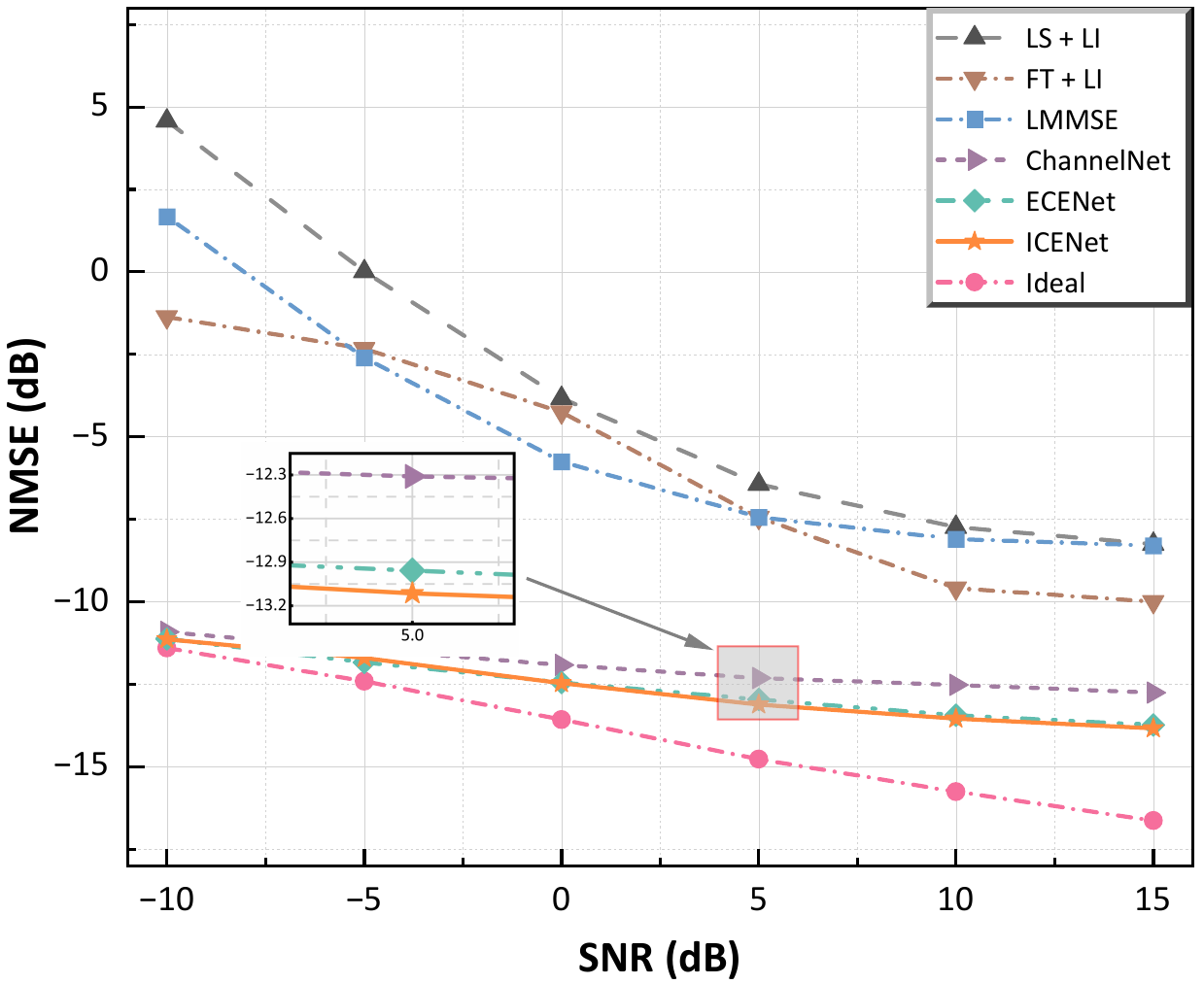}
    \caption{NMSE v.s. SNR for Various Channel Estimation Algorithms with UE at 100km/h.}
    \label{fig: SNRvsNMSE}
\end{figure}

The simulation is conducted for the UMa-NLoS scenario with a center frequency of 3.5GHz and a sounding reference signal interval set to 1ms. Specifically, 14 OFDM symbols are generated each time using the QuaDRiGa and repeated 20000 times to ensure the diversity of each channel slot. Furthermore, the size of the training, validation, and test sets is 15k, 3k, and 2k, respectively.
In terms of network input, the channel dimensions are defined as [$\operatorname{batch}, \operatorname{channel}, \operatorname{subcarrier}, \operatorname{slot}$]. The adaptive moment estimation (Adam) optimizes the model parameters. The initial learning rate, number of epochs, and batch size are set to $10^{-3}$, 100, and 20, respectively. Additionally, a learning rate scheduler with CosineAnnealingLR is utilized, which completes a cosine cycle over 50 epochs, gradually reducing the learning rate to a final value of $10^{-5}$. Moreover, the SNR range for the simulation is set from -10 dB to 15 dB.

The proposed ICENet architecture exhibits performance advantages over other methods, as shown in Fig.~\ref{fig: SNRvsNMSE}. The average number of iterative layers of ICENet is compared with explicit ECENet with a similar number of blocks. For comparison with traditional methods, we utilize the information at the pilot positions and employ the LS method. The Fourier Transform (FT) technique includes power delay profile analysis, noise power estimation and filtering, and then conversion back to the frequency domain. Linearly interpolation (LI) is used to estimate the full channel. The LMMSE algorithm creates the autocorrelation matrix from LS estimates and computes the intercorrelation matrix using LS and LI results (averaged after 500 calculations), followed by average after 50 calculations in the LMMSE formula to finalize the results. In particular, it can be observed from Fig.~\ref{fig: SNRvsNMSE} that the DL-based channel estimation method significantly outperforms conventional techniques in the low signal-to-noise ratio (SNR) region. Moreover, as SNR increases, ICENet requires fewer iterations on average. Furthermore, the ideal lower bound for channel estimation is not practically achievable, as obtaining the true noiseless channel is not feasible in real-world scenarios.

\begin{figure}
    \centering
    \includegraphics[width=0.8\columnwidth]{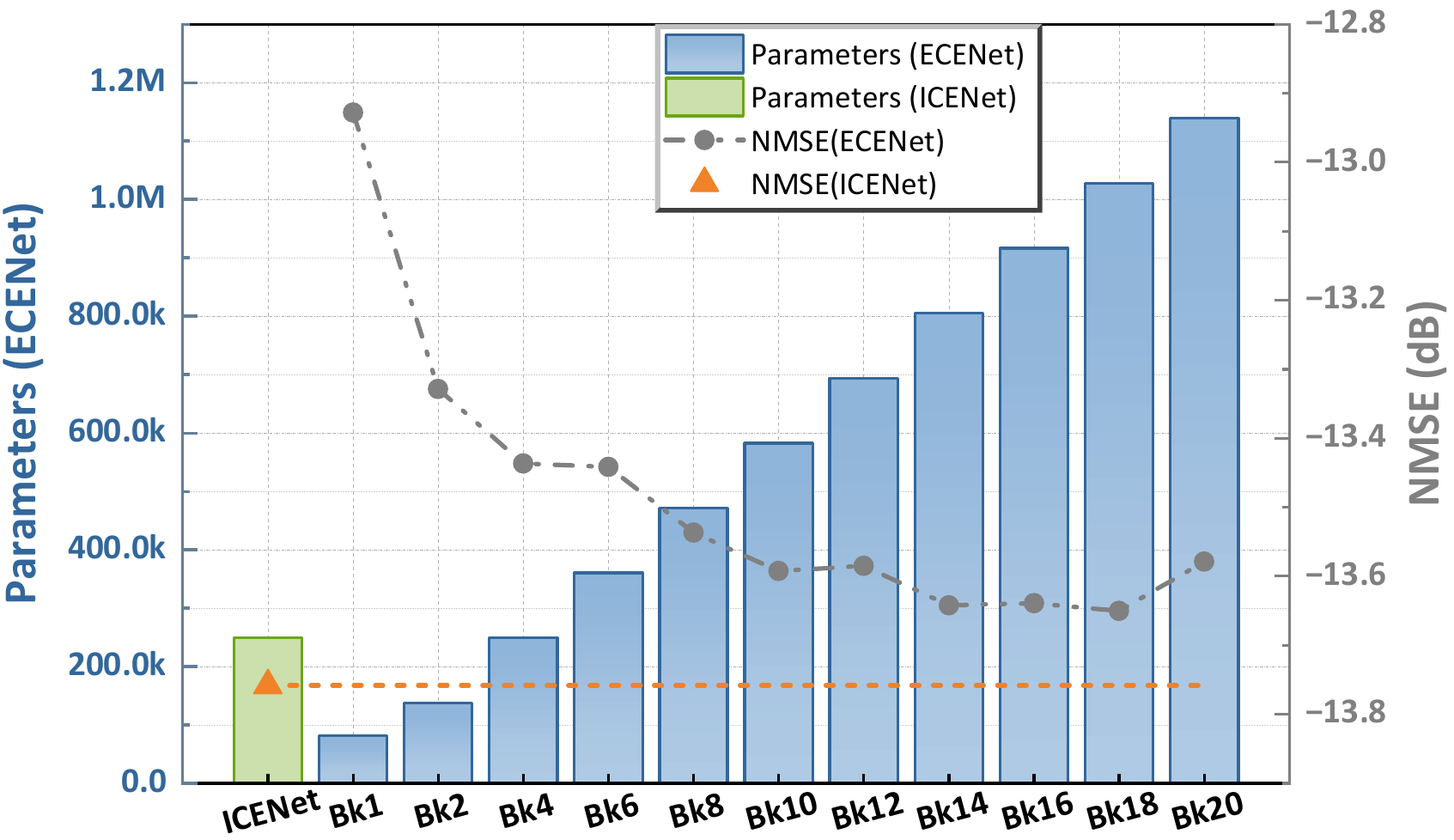}
    \caption{Comparison of the Number of Parameters and NMSE between ICENet and ECENet for UMa-NLoS Scenario with $\text{SNR}$=10dB, $\epsilon$=$10^{-2}$, and $\tau$=10, the UE Speed is 100km/h.}
    \label{fig: ICEvsCE}
\end{figure}

\subsubsection{Adaptive Algorithmic Accuracy and Complexity}
We adopt a lightweight solver, i.e., the AA method. This method accelerates the convergence of the fixed point by utilizing historical iteration information. We set a maximum iteration number in the accelerated iteration process, $\tau$.
Meanwhile, we combine the accuracy requirement, $\epsilon$, of the fixed point solution to measure whether the current solution satisfies the relative error tolerance to realize an early stopping strategy.
Combining $\tau$ and $\epsilon$ allows ICENet to adapt its complexity according to different wireless channel samples.
This method significantly improves computational efficiency and ensures accuracy requirements are met.
It stops the iterations in a timely manner, avoiding unnecessary computational overhead, and thereby optimizing the channel estimation performance. 

In Fig.~\ref{fig: ICEvsCE}, adding more blocks to ECENet initially improves performance, but further increases lead to diminishing returns and potential overfitting. In contrast, ICENet uses fewer parameters, outperforms ECENet, and is better suited for memory-limited environments due to its low memory usage and strong performance. Note that, in this work, we stack 4 lightweight blocks shown in Fig.~\ref{fig: Conv} to form the ICENet-IEB, resulting in the number of iterations multiplied by 4 to make a fair comparison with the ECENet with the corresponding number of blocks.

\begin{figure}
    \centering
    \includegraphics[width=0.6\columnwidth]{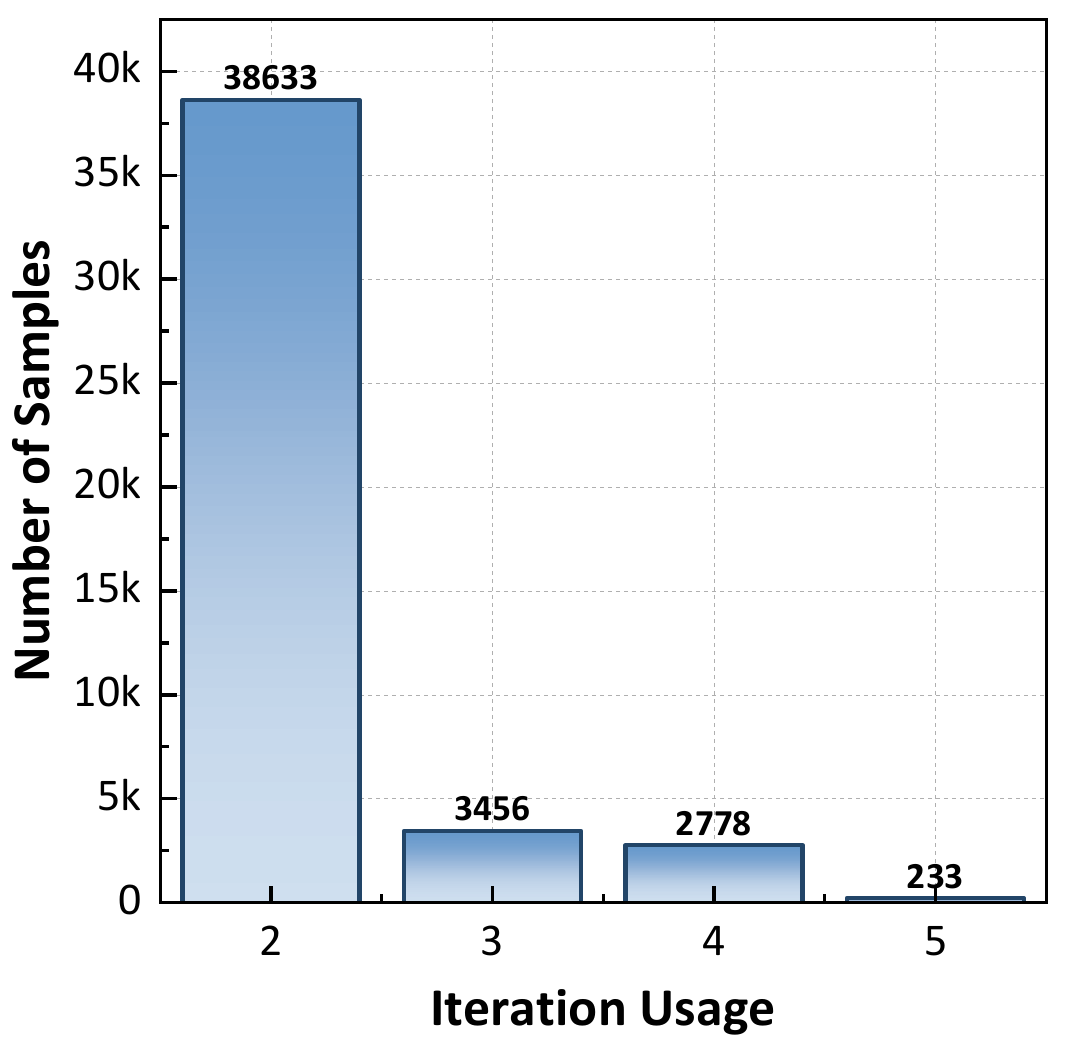}
    \caption{Comparison of the Number of Iterations in ICENet vs. Corresponding Accumulating Number of Channel Samples in the Data Set for UMa-NLoS Scenario with $\text{SNR}$=10dB, $\epsilon$=$1e^{-2}$, and $\tau$=10, the UE Speed is 100km/h.}
    \label{fig: ItervsNum}
\end{figure}

Further, Fig.~\ref{fig: ItervsNum} shows how many channel samples in the data set use the corresponding number of iterations to reach the desired accuracy.
{
The results show that most channel samples are simple inputs, which do not require reaching the maximum iterations $\tau$ as set in the simulation.}
As shown in Fig.~\ref{fig: ICEvsCE}, ICENet has an average of 2.215 iterations (average depth due to varying input needs). Additionally, to achieve better performance, we cascade 4 lightweight blocks in one ICENet-IEB. Therefore, the ICENet is compared with an ECENet stacking with $\lceil2.215\times4\rceil \approx 9$ lightweight blocks. It can be seen that ICENet achieves better performance with fewer layers, compared with the 9-block ECENet, displaying its ability to dynamically adjust the network structure according to the complexity of the samples.
For some channel samples, excessive iterations may lead to performance degradation, while ICENet adaptively reduces unnecessary iterations, thus ensuring superior overall performance.

\begin{table}[t!]
    \centering
    \caption{\textbf{
   ICENet-IEB's Fixed-Point Accuracy vs. NMSE
    }
    }
    \vspace{-1.5em}
    \renewcommand\arraystretch{1}
    \label{tab: ICENet}
    \begin{tabular}{@{}cccc@{}}
        \toprule
        \textbf{Adaptive Standards} & \textbf{IterF-Mean} & \textbf{Parameters} & \textbf{NMSE - Test} \\ \midrule
        $\epsilon$ = 0.5 / $\tau$ = 10 & 2.3 & 81984 & 0.0157 \\
        $\epsilon$ = 0.1 / $\tau$ = 10 & 3.2 & 81984 & 0.0116 \\
        $\epsilon$ = 0.01 / $\tau$ = 20 & 5.6 & 81984 & 0.0106 \\
        $\epsilon$ = 0.001 / $\tau$ = 30 & 9.8 & 81984 & 0.0101 \\ \bottomrule
    \end{tabular}
    \small
    \parbox{\linewidth}{\small \textit{Note:} \scriptsize {This scenario is UMa-NLoS with UE speed of 10km/h and $\text{SNR}$=10dB. IterF-Mean: the average iteration for reaching fixed points in forward propagation. $\epsilon$: the relative error tolerance. $\tau$: the maximum number of iterations.
    }
    }
\end{table}

Apart from the improvements of NMSE, we also find an interesting result for the number of parameters, as seen from Table~\ref{tab: ICENet}, no matter how many iterations we apply in the ICENet network, the number of parameters remains constant.  This is a significant improvement for reducing memory storage, as in explicit DNNs, the number of parameters is usually proportional to the depth of the DNNs. Additionally, it is evident that as the tolerance $\epsilon$ decreases, the number of iterations increases, resulting in a monotonic decrease in the channel estimation NMSE, which emphasizes the adaptive process achieved by our proposed DNN. Note that, as $\epsilon$ decreases, the maximum number of iterations $\tau$ should also be increased because the maximum number of iterations required increases as $\epsilon$ becomes smaller in value.

\section{Open Issues and Research Directions}
The research in \cite{9770094} highlights that increasing the complexity of DL models significantly raises computational demands and costs. ICENet can adapt algorithm complexity to varying channel conditions while maintaining constant and small memory costs, making it suitable for resource-constrained scenarios. However, research on implicit DL architectures in 6G is still in its infancy, with several critical key technical issues open. This section discusses open issues and potential future research directions. 

\subsection{Open Issues}
\begin{itemize}
    \item \textit{Adjustment of Thresholds:}
    In 6G communications, dynamic channel variations demand algorithms capable of adaptive complexity adjustment, a capability provided for ICENet by setting certain thresholds. However, a systematic threshold selection criterion is yet to be established. Future work could integrate communication knowledge to develop information entropy-based threshold quantization and scenario-adaptive threshold strategies.
    
    \item \textit{Stability of Module:}
    Implicit DNNs for 6G must design a stable $f_{\theta}(\cdot)$ capable of converging reliably across diverse channel conditions. Prior studies on implicit DNNs in other fields overlook convergence stability, leading to suboptimal outputs. Future research could apply fixed-point theorems to design targeted regularization strategies based on 6G communication environments.
   
    \item \textit{Choice of Solver:}
    The choice of implicit DNN solvers directly affects computational efficiency. In 6G communications, ultra-low latency constraints require consideration of solver efficiency. Future research could focus on AI-driven acceleration mechanisms and lightweight solver design, including initialization strategies and differentiable hyperparameter optimization.
\end{itemize}

\subsection{Research Directions}
Implicit DL aligns with the 6G architectural vision as a novel AI paradigm. Through parameter-sharing, implicit DL significantly reduces parameter count, making it suitable for communication edge devices with memory and computational constraints. The 6G-oriented evolution of implicit DL can focus on four aspects. Firstly, by developing cross-domain joint implicit frameworks that enable deep coupling of core modules including channel estimation, prediction, feedback, etc. Secondly, systematic exploration should not be limited to semantic information-driven implicit representations, implicit joint optimization of communication rates and radar resolution, or implicit DL extensions for intelligent reflecting surfaces. Thirdly, integrating implicit representations with large models for wireless communication offers a promising avenue. Finally, academia and industry must collaborate across theoretical innovation, hardware co-design, and ecosystem development to advance 6G wireless intelligence.

\section{Conclusion}
This article has presented the ICENet, a framework for channel estimation in 6G wireless networks. By introducing an AI-driven implicit equilibrium DNN, ICENet dynamically balances the complexity and accuracy of channel estimation according to different requirements. Using a lightweight DL network and early stopping strategy, ICENet adjusts its computational resources dynamically. ICENet achieves low and fixed memory usage through a one-block implicit network design. This design significantly reduces memory overhead compared with traditional DL models without sacrificing accuracy, making ICENet ideal for resource-constrained environments. The framework allows flexibility in block design, depending on the specific tasks. Simulation results have shown that ICENet can provide sample-level adaptability while maintaining accuracy. In summary, ICENet offers a novel and promising solution for channel estimation in 6G networks. Its adaptive nature, memory efficiency, and stable performance make it a strong candidate for future research and practical deployment in 6G technology.

\section{Acknowledgment}
The authors Zhen Qiao and Jiang Xue contribute equally to the work. This work is supported by the National Key R\&D Program of  China, No. 2020YFA0713900, the General Program of National Natural Science Foundation of China, No. 62471377, and the Fundamental Research Funds for the Central Universities, No.xzy012025075. The code and data associated with this work can be found in Code Ocean \url{https://codeocean.com/capsule/5565141/tree}.

\section*{Biographies} 
\noindent \textsc{Zhen Qiao} (zhen$\_$@stu.xjtu.edu.cn) received a Master's degree from the University of Birmingham and is pursuing a Ph.D. at Xi'an Jiaotong University, China. His research interests are artificial intelligence and wireless communications.
\\

\noindent \textsc{Jiang Xue} [S'09, M'13, SM'19] (x.jiang@xjtu.edu.cn) is a Full Professor with School of Mathematics and Statistics, Xi’an Jiaotong University, also with Pengcheng Lab and Pazhou Laboratory (Huangpu), China. His main interests include machine learning and wireless communication.
\\

\noindent \textsc{Junkai Zhang} [S'21, M'24] (jk.zhang@xjtu.edu.cn) is an Assistant Professor with School of Mathematics and Statistics, Xi'an Jiaotong University, China. His research interests include 6G AI-based wireless communications, full duplex radio, and integrate sensing and communications. 
\\

\noindent \textsc{Guanzhang Liu} (lgzh97@stu.xjtu.edu.cn) is pursuing a Ph.D. at Xi’an Jiaotong University, China. His current research interests include DL for intelligent wireless communications, and channel prediction in high mobility scenarios.
\\

\noindent \textsc{Xiaoqin Ma} (mlshr424242@stu.xjtu.edu.cn) is currently a Master's student at Xi’an Jiaotong University, China. Her research interests unified architecture for intelligent PHY.
\\

\noindent \textsc{Runhua Li} (lirh20150802@stu.xjtu.edu.cn) is currently pursuing a Ph.D. degree at Xi’an Jiaotong University, China. His current research interests include DL for PHY and massive MIMO.
\\

\noindent \textsc{Faheem A. Khan} [M'02] (F.Khan@hud.ac.uk) is currently a Reader with School of Computing and Engineering, University of Huddersfield, U.K. His research interests include machine learning and DL in wireless communications, and IoT networks.
\\

\noindent \textsc{John S. Thompson} [M’03, SM’12, F’16] (John.Thompson@ed.ac.uk) is the Personal Chair of Signal Processing and Communications with School of Engineering, The University of Edinburgh, U.K. He specializes in energy-efficient wireless communications and the application of machine learning to wireless communications.
\\

\noindent \textsc{Zongben Xu} (zbxu@xjtu.edu.cn) now serves as the Chief Scientist of the National Basic Research Program of China and Director of the Institute for Information and System Sciences of Xi'an Jiaotong University, China. He was elected as an academician of the Chinese Academy of Science in 2011. His current research interests include intelligent information processing and applied mathematics.


\begin{thebibliography}{1}
\bibliographystyle{IEEEtran}

\bibitem{9770094}
W. Tong and G. Y. Li, ``Nine Challenges in Artificial Intelligence and Wireless Communications for 6G," \textit{IEEE Wireless Commun. Mag.}, vol. 29, no. 4, pp. 140-145, Aug. 2022.

\bibitem{9923616}
W. Xu \textit{et al.}, ``Computer Vision Aided mmWave Beam Alignment in V2X Communications," \textit{IEEE Trans. Wireless Commun.}, vol. 22, no. 4, pp. 2699–2714, April 2023.

\bibitem{9786750}
C. Wang \textit{et al.}, ``Pervasive Wireless Channel Modeling Theory and Applications to 6G GBSMs for All Frequency Bands and All Scenarios," \textit{IEEE Trans. Veh. Technol.}, vol. 71, no. 9, pp. 9159–9173, Sep. 2022.

\bibitem{9023918}
C. Wang \textit{et al.}, ``Artificial Intelligence Enabled Wireless Networking for 5G and Beyond: Recent Advances and Future Challenges," \textit{IEEE Wireless Commun. Mag.}, vol. 27, no. 1, pp. 16–23, Feb. 2020.

\bibitem{BEM}
B. Wang \textit{et al.}, ``Spatial- and Frequency-Wideband Effects in Millimeter-Wave Massive MIMO Systems," \textit{IEEE Trans. Signal Processing}, vol. 66, no. 13, pp. 3393–3406, May 2018.

\bibitem{8715338}
H. He \textit{et al.}, ``Model-Driven Deep Learning for Physical Layer Communications," \textit{IEEE Wireless Commun. Mag.}, vol. 26, no. 5, pp. 77–83, Oct. 2019.

\bibitem{hu2020deep}
Q. Hu \textit{et al.}, ``Deep Learning for Channel Estimation: Interpretation, Performance, and Comparison," \textit{IEEE Trans. Wireless Commun.}, vol. 20, no. 4, pp. 2398–2412, April 2020.

\bibitem{10075639}
J. Gao \textit{et al.}, ``Deep Learning-Based Channel Estimation for Wideband Hybrid MmWave Massive MIMO," \textit{IEEE Trans. Commun.}, vol. 71, no. 6, pp. 3679–3693, June 2023.

\bibitem{soltani2019deep}
M. Soltani et al., ``Deep Learning-Based Channel Estimation," \textit{IEEE Commun. Lett.}, vol. 23, no. 4, pp. 652–655, April 2019.

\bibitem{bai2022equilibrium}
S. Bai, ``Equilibrium Approaches to Modern Deep Learning," April 2022; \url{https://kilthub.cmu.edu/articles/thesis/Equilibrium_Approaches_to_Modern_Deep_Learning/21580017}, accessed on Jan. 30, 2024.

\bibitem{quadriga}
S. Jaeckel \textit{et al.}, ``QuaDRiGa - Quasi Deterministic Radio Channel Generator, User Manual and Documentation," \textit{Fraunhofer Heinrich Hertz Institute, Tech. Rep. v2.8.1}, Dec. 2023.

\bibitem{xusun}
Z. Xu and J. Sun, ``Model-Driven Deep-Learning," \textit{National Science Review}, vol. 5, no. 1, pp. 22–24, Aug. 2017.

\bibitem{zhang2023dive}
A. Zhang \textit{et al.}, \textit{Dive into Deep Learning}. Cambridge University Press, Feb. 2023.

\bibitem{10310231}
P. Jiang \textit{et al.}, ``Dual CNN-Based Channel Estimation for MIMO-OFDM Systems," \textit{IEEE Trans. Commun.}, vol. 69, no. 9, pp. 5859–5872, Sep. 2021.

\bibitem{anderson1965iterative}
D. G. Anderson, ``Iterative Procedures for Nonlinear Integral Equations,"
\textit{Journal of the ACM}, vol. 12, no. 4, pp. 547–560, Oct. 1965.

\end{thebibliography}
\end{document}